\documentclass[fleqn,usenatbib]{mnras}

\usepackage{newtxtext,newtxmath}

\usepackage[T1]{fontenc}
\usepackage{ae,aecompl}

\usepackage{graphicx}	
\usepackage{amsmath}	
\usepackage{amssymb}	
\usepackage{bm}

\title[Multi-planet-disc interactions]{Multi--planet disc interactions in binary systems}

\author[A. Franchini et al.]{
Alessia Franchini$^{1}$\thanks{E-mail: alessia.franchini@unlv.edu},
Rebecca G. Martin$^{1}$ and
Stephen H. Lubow$^{2}$
\\
$^{1}$Department of Physics and Astronomy, University of Nevada, 4505 South Maryland Parkway, Las Vegas, NV 89154, USA\\
$^{2}$Space Telescope Science Institute, 3700 San Martin Drive, Baltimore, MD 21218, USA\\
}

\date{Accepted 2019 October 24. Received 2019 October 6; in original form 2019 June 12.}

\pubyear{2019}
\begin{document}
\label{firstpage}
\pagerange{\pageref{firstpage}--\pageref{lastpage}}
\maketitle

\begin{abstract}
We investigate the evolution of a multi--planet--disc system orbiting one component of a binary star system. The planet--disc system is initially coplanar but misaligned to the binary orbital plane. The planets are assumed to be giants that open gaps in the disc. We first study the role of the disc in shaping the mutual evolution of the two planets using a secular model for low initial tilt. In general we find that the planets and the disc do not remain coplanar, in agreement with \cite{Lubow2016} for the single planet case. Instead, the planets and the disc undergo tilt oscillations. A high mass disc between the two planets causes the planets and the disc to nodally precess at the same average rate but they are generally misaligned. The amplitude of the tilt oscillations between the planets is larger while the disc is present. We then consider higher initial tilts using hydrodynamical simulations and explore the possibility of the formation of eccentric Kozai-Lidov (KL) planets. We find that the inner planet's orbit undergoes  eccentricity growth for a large range of disc masses  and initial misalignments.  For a low disc mass and large initial misalignment both planets and the disc can undergo KL oscillations. Furthermore, we find that sufficiently massive discs can cause the inner planet to increase its inclination beyond $90^{\circ}$ and therefore to orbit the binary in a retrograde fashion. The results have important implications for the explanation of very eccentric planets and retrograde planets observed in multi-planet systems.

\end{abstract}




\section{Introduction}

The discovery of thousands of exoplanets presents a variety of exotic configurations of their orbits around the host star. 
\cite{Horch2014,Deacon2016,Matson2018,Ziegler2018} estimated that roughly 50\% of these planets might be hosted by binary systems.
Planet formation in a binary system is particularly worth investigating in order to explain high inclination, eccentric close-in planets that have been observed.

A large fraction of observed exoplanet orbits are highly misaligned with respect to the spin of the host star \citep{Triaud2010,Winn2010,Winn2015,Triaud2018}. The misalignment is measured with the Rossiter-McLaughlin effect \citep{Rossiter1924,McLaughlin1924}. There are three main mechanisms that have been proposed to explain the observed misalignments. First, a planet might lie on a misaligned orbit as a result of the interaction with another object, either a star or a planet \citep[e.g][]{Takeda2008}. Second, the misalignment might arise because the protoplanetary disc feels a torque exerted by a binary companion \citep[e.g.][]{Batygin2012}.  Finally, the magnetic field of the star may induce changes in the orientation of the spin of the star relative to the disc angular momentum \citep{Lai2011}. In this work we consider the second effect.

Among the planets that show projected spin-orbit misalignments, a significant fraction show apparent retrograde motion \citep{Triaud2010}. For instance, observations of the Rossiter-McLaughlin effect confirmed the prediction made by \cite{Anderson2010} on the nature of the retrograde orbit of WASP-17b.

According to the current commonly accepted scenario, planets form in a protoplanetary disc. 
Giant planets are believed to be embedded in the disc around the host star in the earliest stages of their formation \citep{Pollack1996,Hubickyj2005,Papaloizou2005}.
When the planet grows to the mass of Neptune, it undergoes runaway-gas accretion and its tidal torque opens a gap in the gas disc \citep{lin1986,Dangelo2002,Bate2003}.
In this scenario, these planets form outside the snow line  because the presence of ice enhances the surface density of the planetesimals \citep{Lecar2006}. They are also required to form within the lifetime of the gas disc, roughly a few million years \citep{Haisch2001,Pascucci2006,Fedele2010}.
Terrestrial planets are believed to form from agglomeration of Moon to Mars sized planetary embryos and planetesimals \citep{Chambers2001,Obrien2006}. Therefore, while these kind of small planets are most likely to remain coupled with the gaseous disc, giant planets interact more strongly with the disc. The study of these kind of interactions is essential in order to explain the presence of giant planets on inclined and/or eccentric orbits close to the host star.

Understanding this kind of system is key for explaining observations such as the Kepler-56 planetary system \citep{Huber2013}. This is a red giant star hosting two close-in planets of Neptune and Saturn size respectively on highly inclined orbits with respect to the spin of the star. Misalignments between the planet orbits and the host star spin are believed to be caused by the torque exerted from a wide orbiting companion star. Radial velocity measurements revealed indeed a third object in Kepler-56.

Once a planet is massive enough to form a gap in the gas disc, the binary torque dominates that of the disc. \cite{Picogna2015}, \cite{Lubow2016} and \cite{Martin2016} found that it is more likely for the planet and the disc to undergo inclination oscillations on long timescales and not to remain coplanar. The phase angles of the components that orbit the primary star can either be in a librating or circulating state. In the librating state, the phase difference between different components is limited.  Therefore the mean precession rates for the librating objects are locked and they evolve together.
If instead the phase angles have different mean precession rates,  the system is circulating and the components evolve independently from one another.

In this paper we extend the work done by \cite{Martin2016} and \cite{Lubow2016} on a single planet--disc system to investigate multi-planet systems. We study the orbital evolution of two giant planets with a gas disc.
In Section~\ref{sec:script} we first use analytical calculations to investigate the role that the accretion disc has in shaping the planets orbital evolution for low initial misaligment.
Then in Section~\ref{sec:sim} we use smoothed particle hydrodynamical (SPH) simulations to explore higher initial misalignment and the possibility of planets undergoing Kozai-Lidov oscillations \citep{Kozai1962,Lidov1962}.
In principle, if the two planets are circulating, the inner planet is likely to undergo KL oscillations if its inclination exceeds the critical value $i_{\rm crit}=39^{\circ}$. On the contrary, if the distance between the planets is small enough, there is a chance that the outer planet suppresses the KL mechanism acting on the inner planet.
From the results presented in \cite{Lubow2016} it is clear that the disc mass is a crucial parameter to determine whether the planets evolve together or separately.
We therefore investigate in this paper the role of the disc mass in shaping the planets evolution and determining the possibility of KL oscillations of the inner planet. We draw our conclusions in Section~\ref{sec:concl}.

\begin{figure*}
    \includegraphics[width=\columnwidth]{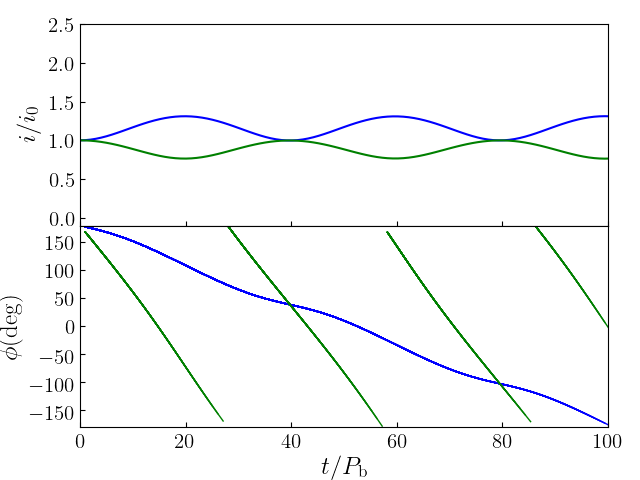}    \includegraphics[width=\columnwidth]{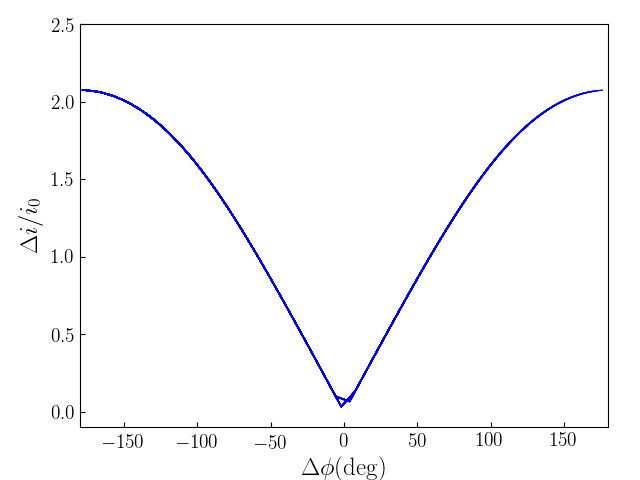}
    \caption{Left panel: secular tilt (upper) relative to the binary orbital plane and phase (lower) evolution of a planet-planet system that orbits around one component of a circular binary. The time is in units of the orbital period of the binary $P_{\rm b}$. Initially the two planets are coplanar and lie on a misaligned orbit with inclination $i_0$ with respect to the binary orbital plane. The blue line represents the inner planet while the green line is the outer planet. 
    Right panel: Phase portrait of the relative planet-planet tilt $\Delta i/i_0$ vs. nodal phase difference $\Delta \phi$.}
    \label{fig:nodiscs}
\end{figure*}

\begin{figure*}
    \includegraphics[width=\columnwidth]{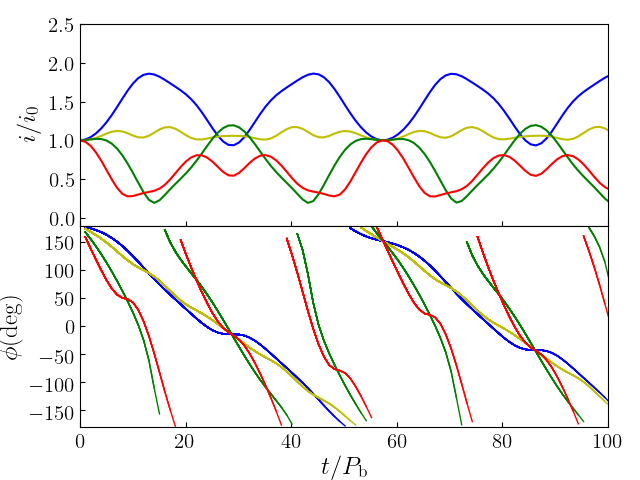}    \includegraphics[width=\columnwidth]{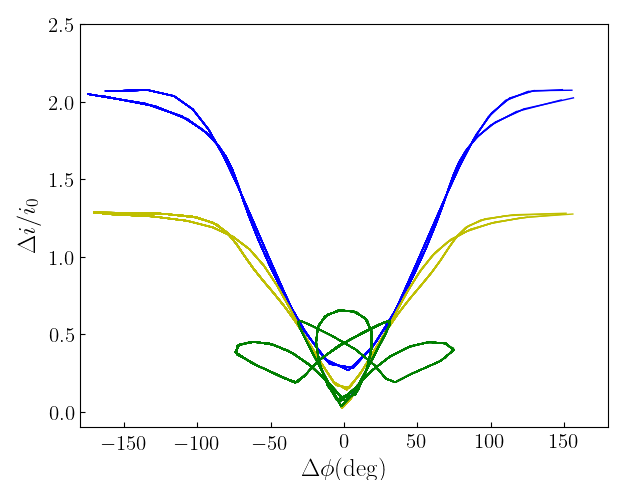}
    \caption{Left panel: secular tilt relative to the binary orbital plane (upper) and phase (lower) evolution of a planet-disc-planet-disc system that orbits around one component of a circular binary. The time is in units of the orbital period of the binary $P_{\rm b}$. Initially the two planets and the two discs are coplanar and lie on a misaligned orbit with respect to the binary orbital plane. The blue, yellow, green and red lines represents the inner planet, inner disc, outer planet and outer disc respectively. The total mass of the disc is $M_{\rm d}=0.001\,M$.
    Right panel: Phase portrait of the relative planet-planet (blue), inner disc-outer planet (yellow) and outer planet-outer disc (green) tilt $\Delta i/i_0$ vs. nodal phase difference $\Delta \phi$.}
    \label{fig:discs_m0.001}
\end{figure*}

\begin{figure*}
    \includegraphics[width=\columnwidth]{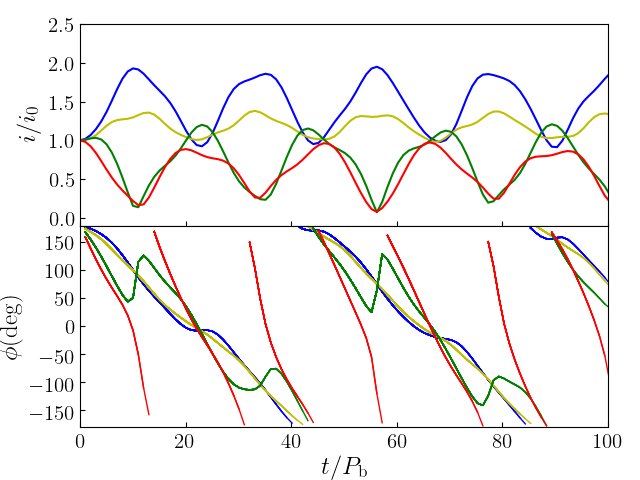}    \includegraphics[width=\columnwidth]{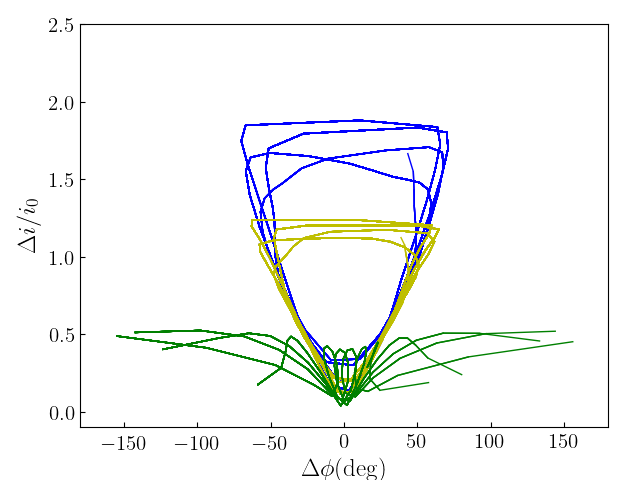}   
    \caption{Same as Figure \ref{fig:discs_m0.001} but with total disc mass $M_{\rm d}=0.002\,M$.}
    \label{fig:discs_m0.002}
\end{figure*}

\section{Analytical calculations: secular evolution}
\label{sec:script}

The system under investigation is composed of a primary star surrounded by a gas disc with two planets embedded in it and a companion star orbiting the primary.
In the secular analytical calculations we assume that both planets already carved a gap in the disc so that the objects orbiting the primary are in the following order: inner planet, inner disc, outer planet, outer disc and companion star.

The six bodies interact only via gravitational forces, the discs are taken to be non-viscous, rigid and flat. Since there is no dissipation within the disc, the amplitude of the inclination oscillations between the discs and the planets remains constant with time in the case where there is only one planet interacting with one disc. In the multi--planet--disc system the oscillations become more chaotic.
The orbital planes of the planets and the two discs are initially coplanar but misaligned with respect to the binary orbital plane.

The binary is assumed to be circular, equal mass with total mass $M=M_1+M_2$ and separation $a$. We put two planets with mass $M_{\rm p,i}=0.001\,M$ around the primary star located at orbital radii $d_1=0.05\,a$ and $d_2=0.15\,a$ respectively.

We adopt the same procedure outlined in \cite{Lubow2016} describing the time dependent tilt of each object relative to the binary plane with the complex variable $W(t) = l_{\rm x}(t) + il_{\rm y}(t)$, where the unit angular momentum vector as a function of time is $\bm{l}(t)=(l_{\rm x}(t),l_{\rm y}(t),l_{\rm z}(t))$. 
Assuming that the tilts are small, we solve the secular evolution equations for the gravitational interaction between the different objects in the system \citep{Lubow2001}. 
According to this model, the torques are evaluated in the small angle approximation. As \cite{Lubow2016} pointed out, the analytical results are therefore accurate for initial tilts up to about $20^{\circ}$.
Furthermore this approach assumes the orbits to remain circular, therefore it cannot be used to study systems undergoing KL oscillations.
We investigate higher inclination systems using SPH simulations in Section \ref{sec:sim}.

The interaction between the components of the system is described by a linear model with coupling coefficients.
The evolution equations for the planets tilt $W_{\rm p1},W_{\rm p2}$ and the two discs $W_{\rm d1},W_{\rm d2}$ are 
\begin{align}
\begin{split}\label{eq:1}
    J_{\rm p1} \frac{dW_{\rm p1}}{dt} = {}& iC_{\rm p1d1}(W_{\rm d1} - W_{\rm p1}) + iC_{\rm p1p2}(W_{\rm p2} - W_{\rm p1}) \\
    & + iC_{\rm p1d2}(W_{\rm d2} - W_{\rm p1}) - iC_{\rm p1s}W_{\rm p1}
\end{split}\\
\begin{split}\label{eq:2}
    J_{\rm p2} \frac{dW_{\rm p2}}{dt} = {}& iC_{\rm p2d1}(W_{\rm d1} - W_{\rm p2}) + iC_{\rm p1p2}(W_{\rm p1} - W_{\rm p2}) \\
    & + iC_{\rm p2d2}(W_{\rm d2} - W_{\rm p2}) - iC_{\rm p2s}W_{\rm p2}
\end{split}\\
\begin{split}\label{eq:3}
    J_{\rm d1} \frac{dW_{\rm d1}}{dt} = {}& iC_{\rm p1d1}(W_{\rm p1} - W_{\rm d1}) + iC_{\rm d1p2}(W_{\rm p2} - W_{\rm d1}) \\
    & + iC_{\rm d1d2}(W_{\rm d2} - W_{\rm d1}) - iC_{\rm d1s}W_{\rm d1}
\end{split}\\
\begin{split}\label{eq:4}
    J_{\rm d2} \frac{dW_{\rm d2}}{dt} = {}& iC_{\rm d2d1}(W_{\rm d1} - W_{\rm d2}) + iC_{\rm d2p2}(W_{\rm p2} - W_{\rm d2}) \\
    & + iC_{\rm d1d2}(W_{\rm d1} - W_{\rm d2}) - iC_{\rm d2s}W_{\rm d2}
\end{split}
\end{align}
where p1, d1, p2, d2 and s denote respectively the inner planet, inner disc, outer planet, outer disc and companion star with $J$ being the respective angular momentum. 
The coupling coefficients describing the interaction of the various components of the system have the same form as the ones in \cite{Lubow2016}.

In order to understand the role of the disc in this type of architecture, we first consider only the evolution of the two planets misaligned with respect to the binary orbital plane, i.e. we ignore the presence of the disc.
The coupling coefficient between the planets is given by
\begin{equation}
    C_{\rm p1p2} = G\,M_{\rm p1}M_{\rm p2}\,K(d_1,d_2)
    \label{eq:ppcoeff}
\end{equation}
and the interaction between the planets and the companion star has the same form as Eq. \ref{eq:ppcoeff} (see Eq. 7 in \cite{Lubow2016}). The kernel $K(d_1,d_2)$ has the same form as in \cite{Lubow2016}. 
The left panel of Figure \ref{fig:nodiscs} shows the inclination and phase angles evolution of the two planets.
The two objects oscillate away from each other and from the phase angle evolution we can see that each planet precesses at its own rate.
Looking at the phase portrait in the right panel of Figure \ref{fig:nodiscs}, we see that the planets are circulating, meaning that they are evolving independently from one another. 
The inclination of each component relative to the binary orbital plane is defined as $i_{\rm j}(t)=|W_{\rm j}(t)|$ and the phase angle is $\phi_{\rm j}(t)=\tan^{-1}(\Im[W_{\rm j}(t)]/\Re[W_{\rm j}(t)])$ where $j=\,\rm p1,\, p2,\, d1,\, d2,\, s$. 
The relative tilt between the $i,j$ components is defined as $\Delta i/i_0=|W_{\rm i}-W_{\rm j}|$ while the nodal phase difference is $\Delta \phi = \phi_{\rm i}-\phi_{\rm j}$.

We then add the two portions of the accretion disc in the calculations and investigate their effect on the planets evolution. We assume that the mass contained in each portion of the disc is the same and the discs extend from $R_{\rm in,1}=0.06\,a$ to $R_{\rm out,1}=0.125\,a$ and from $R_{\rm in,2}=0.175\,a$ to $R_{\rm out,2}=0.25\,a$ respectively. The discs aspect ratio at the inner edge is $H/R(R_{\rm in,i})=0.035$, the sound speed has the form $c_{\rm s}\propto (R/R_{\rm in,i})^{-q}$ with $q=0.75$, the surface density profile in the two discs is $\Sigma_1 \propto (R/R_{\rm in,1})^{-p}$ and $\Sigma_2 \propto (R/R_{\rm in,2})^{-p}$ with $p=1.5$. The index $i=1,2$ indicates the two discs. The constant of proportionality for the surface densities is determined by the mass contained in each disc, which is half of the total disc mass.  
The interaction between the planets and the discs is described as
\begin{equation}
    C_{{\rm p}i{\rm d}j} = 2\pi \int_{R_{{\rm in},j}}^{R_{{\rm out},j}} G\,M_{{\rm p}i}\,R \Sigma_j(R)\,K(R,d_{\rm i})\,dR,
    \label{eq:pdcoeff}
\end{equation}
where $i=1,2$ indicates the inner and outer planet and $j=1,2$ indicates the inner and outer disc respectively.
We also introduce a new coupling coefficient that describes the interaction between the inner and outer disc in the form
\begin{equation}
   C_{\rm d1d2} =  (2\pi)^2 \int_{R_{\rm in,2}}^{R_{\rm out,2}} \int_{R_{\rm in,1}}^{R_{\rm out,1}} r_1\, r_2\,\Sigma_1(r_1)\,\Sigma_2(r_2)\, K(r_1,r_2)\,dr_1\,dr_2
   \label{eq:ddcoeff}
\end{equation}
where $x$ represent the radius inside the inner disc and $y$ is the radius within the outer disc. 

Figure \ref{fig:discs_m0.001} shows the results for the same system but including the inner and outer discs. The total disc mass is $M_{\rm d}=0.001\,M$.
The upper left panel of Fig. \ref{fig:discs_m0.001} shows the evolution of the inclination of the inner (blue) and outer (green) planet and of the inner (yellow) and outer (red) disc.
The lower left panel of Fig. \ref{fig:discs_m0.001} shows that the two planets (blue and green line) precess independently.
The right panel of Figure \ref{fig:discs_m0.001} shows the phase portrait of the relative planet-planet (blue), inner disc-outer planet (yellow) and outer planet-outer disc (green) tilt. We see that the inner planet is circulating with the outer planet but it is librating with the inner disc (see blue and yellow lines in the lower left panel).  The outer planet is librating with the outer disc (green line in the phase portrait). 
Therefore, a low mass disc with a total mass of $M_{\rm d}=0.001\,M$ leaves the planets free to circulate and precess each one with its own frequency around the primary star. However, the amplitude of the tilt oscillations relative to the binary orbital plane are significantly increased by the presence of the disc. 

Figure \ref{fig:discs_m0.002} shows the results for the same system but with a higher total disc mass $M_{\rm d}=0.002\,M$.
The lower left panel of Fig. \ref{fig:discs_m0.002} shows that the two planets (blue and green line) precess at the same rate together with the portion of the disc between them (yellow line).
The outer disc (red line) behaves independently with respect to the three inner components. However, the outer planet librates with the outer disc for very brief intervals during the evolution of the system, as seen in the closed green lines in the  right panel of Figure \ref{fig:discs_m0.002}.
Also in that panel, we see that the inner planet is librating  with the outer planet (blue line) and the outer planet librates with the inner disc (yellow line).
Therefore, increasing the disc mass results in librating solutions of the two planets and the portion of the disc between them. For a higher mass disc the relative peak inclination drops (blue line in right panel of \ref{fig:discs_m0.002}),  as expected from the action of the secular resonance (see Figure 2 in \cite{Lubow2016}). 
The amplitude of the tilt oscillations of the planets relative to the binary orbital plane are again much larger as a result of the presence of the disc.

We see from Figs. \ref{fig:nodiscs}-\ref{fig:discs_m0.002} that the presence of the two discs leads also to an increase in the mutual inclination between the planets. 
When the disc is present, the inclination of the inner planet relative to the binary orbital plane increases by roughly a factor two with respect to the initial inclination. 
Therefore, even if the same system starts with a $i_0=20^{\circ}$ inclination, it is likely that the planet reaches an inclination above the critical angle for the KL mechanism to operate.
Since the secular model is valid for small tilts (i.e. up to $20^{\circ}$) we need to explore the higher inclination regime with hydrodynamical simulations.

\begin{figure}
    \centering
    \includegraphics[width=\columnwidth]{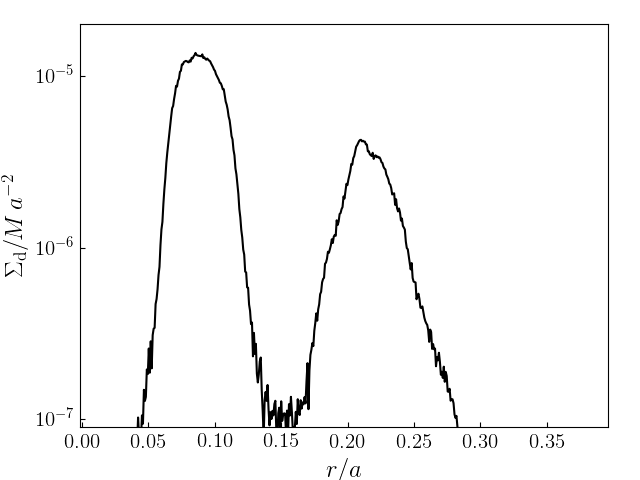}
    \caption{Surface density profile in log scale at a time of 4 binary orbits for a coplanar disc with initial mass of $M_{\rm d,in}=4\times 10^{-6}M$. The surface density $\Sigma$ is normalized by $M/a^{2}$, the binary mass divided by the square of its separation. The simulations start with the two planets embedded in the disc without a gap, as discussed in Section \ref{sec:setup}.}
    \label{fig:init_sig}
\end{figure}

\section{Multi-planet disc hydrodynamical simulations}
\label{sec:sim}

In order to further study the behaviour of the planets embedded in the accretion disc, including the viscous evolution of the disc and higher initial inclinations, we perform Smoothed Particle Hydrodynamics (SPH) numerical simulations using the code {\sc phantom} \citep{Lodato2010,Price2010,Price2012,Price2017,Nixon2012,Nixon2013}.  Planets embedded in accretion discs around one component of a binary star system have been previously studied with {\sc phantom} \citep[e.g.][]{Lubow2016,Martin2016}. 
We perform SPH simulations using $N=5 \times 10^5$ particles initially. 
The resolution of the simulation depends on $N$, the viscosity parameter $\alpha$ and the disc scale height $H$. 
The \cite{SS1973} viscosity parameter is modelled by adapting the artificial viscosity according to the approach of \cite{Lodato2010}. 
The main implication of this treatment is that in order to provide a constant $\alpha$ in the disc, the disc scale height $H$ must be uniformly resolved.
This is achieved by choosing both a surface density profile and a temperature profile that ensures this condition \citep{Lodato2007}, as discussed below. 

Disc evolution strongly depends on the value of the protoplanetary discs viscosity which is not well known.
In our simulations, we model disc viscosity using the artificial viscosity parameter $\alpha_{\rm AV}$ \citep{Lodato2010}.
The main consequence of modelling disc viscosity through the artificial viscosity parameter is that there is a lower limit below which a physical viscosity is not resolved in SPH ($\alpha_{\rm AV}\approx 0.1$). Viscosities smaller than this value can produce disc spreading that is independent of the value of $\alpha_{\rm AV}$ \citep{Bate1995,Meru2012}.

\subsection{Simulation set--up}
\label{sec:setup}

If we simply place the formed planets into the disc (with no gap at their orbital location), the two planets and the disc can lose the initial coplanarity 
before the two gaps are carved.   Furthermore, the planets can increase their mass significantly before carving the gaps and we would lose control over the mass of the planet when an equilibrium gap is established. Thus, we follow the methods described in \cite{Lubow2016} and find the surface density of a disk with an equilibrium gap already established at the start of the simulation.

We consider an initial configuration with two planets embedded in an accretion disc that orbits the primary star. Both planets and the disc are coplanar and lie in the same plane as the binary. The equal mass binary has total mass $M=M_1 + M_2=1$ and a circular orbit in the $x$-$y$ plane with separation $a$.
We put two planets with mass $M_{\rm p}=0.001\,M$ around the primary star located at $d_1=0.05\,a$ and $d_2=0.15\,a$ respectively. 
The two stars and the two planets are all modelled as sink particles. We choose the accretion radius for the particle removal to be $R_{\rm acc}=0.0125\,a$ for the stars and $10$\% of the Hill radius around each planet ($0.0044\,a$ and $0.0013\,a$ respectively for the inner and outer planet). 
The simulation results do not vary significantly for smaller values of these accretion radii. The mass and angular momentum of any particle that enters the sink radius are added to that of the sink particle.

We start with a very low initial disc mass of $M_{\rm d,in}=4\times 10^{-6}M$ and an initial disc radial extent from $R_{\rm in}=0.0125\,a$ to $R_{\rm out}=0.225\,a$. 
The disc then slightly expands outwards reaching $R_{\rm out}=0.25\,a$ which corresponds to the tidal truncation radius of a disc in a coplanar binary system with separation $a$ \citep{Paczynski1977}. 

The initial surface density profile of the disc is set as $\Sigma\propto (R/R_{\rm in})^{-3/2}(1-\sqrt{R_{\rm in}/R})$ between $R_{\rm in}=0.0125\,a$ and $R_{\rm out}=0.225\,a$.
With this density distribution and a sound speed given by $c_{\rm s}\propto (R/R_{\rm in,i})^{-q}$ with $q=0.75$, the disc is uniformly resolved. The circumprimary disc aspect ratio at the inner edge is chosen to be $H/R(R_{\rm in})=0.035$ which is a typical value for protoplanetary discs. The viscosity parameter is chosen to be $\alpha=0.01$.

The system evolves in the plane of the binary (i.e. $i=0^{\circ}$) until the planets carve two gaps in the disc.
We find that a reasonably stable equilibrium gap structure is established in roughly $50$ outer planet orbits, or 4 binary orbital periods. Figure \ref{fig:init_sig} shows the disc surface density profile after 4 binary orbits. The portion of the disc that lies inside the inner planet is accreted onto the star so the resulting system orbiting the primary star is composed of an inner planet, a disc, an outer planet and an outer disc. 
The amount of mass contained in each disc is roughly half of the total disc mass. 

\begin{figure*}
    \centering
    \includegraphics[width=2\columnwidth]{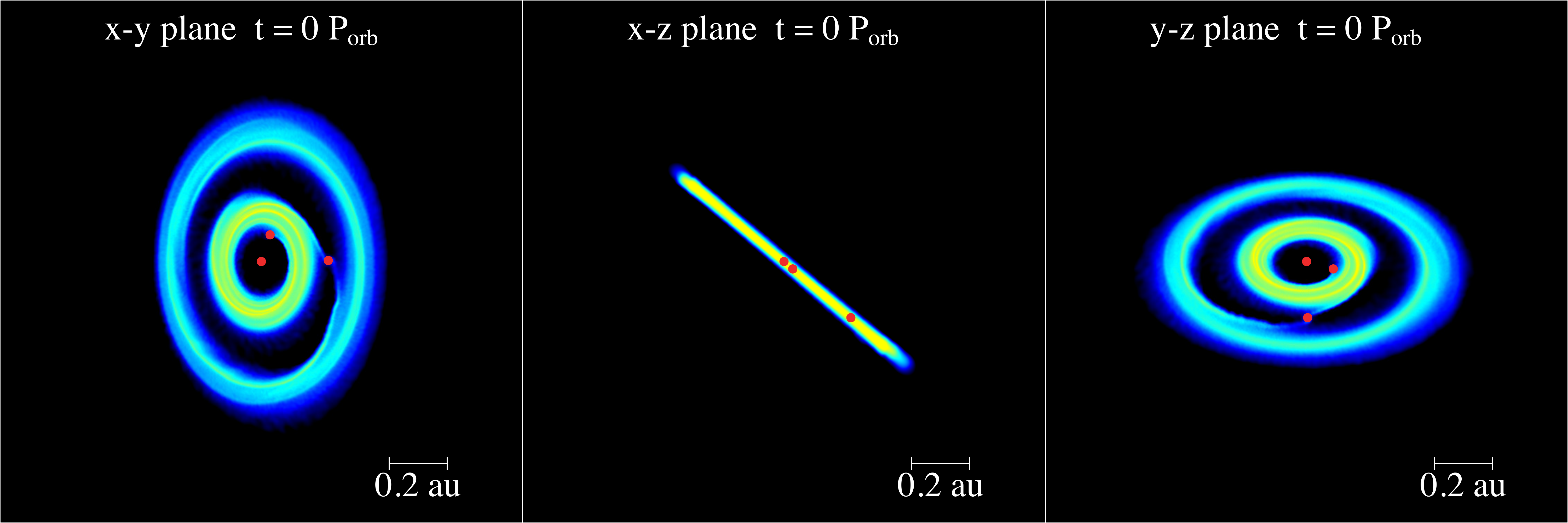}
     \includegraphics[width=2\columnwidth]{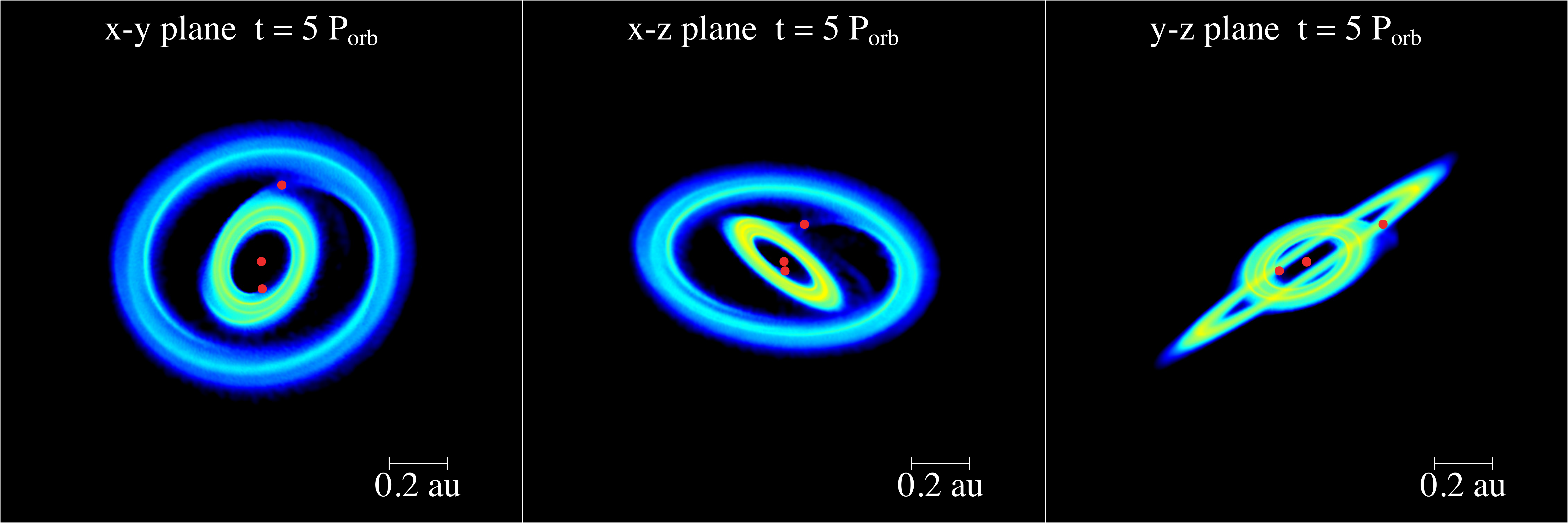}
    \caption{ Simulation with initial disc mass $M_{\rm d}=0.002\,M$ and tilt $40^\circ$. The colour of the gas denotes the column density with yellow being about two orders of magnitude larger than blue. The red points show the primary star (located at the centre of each panel) and the two planets. The binary companion is not shown on this scale but is located along the positive $x$-axis. The left panels show the view looking down on the binary orbital plane, the $x-y$ plane, the middle panels show the $x-z$ plane and the right panels show the $y-z$ plane. The upper panels show the initial conditions and the lower panels show the system after a time of $5\,P_{\rm orb}$. }  
    \label{fig:splash}
\end{figure*}

The disc density profile shape does not depend on the disc mass if the disc is locally isothermal and non-selfgravitating and if the planets are on fixed orbits. Note that planet migration occurs on a much longer timescale than the timescales considered in this work. 
Therefore, in the rest of our simulations, we rescale this density distribution obtained after $4$ binary orbital periods to achieve the desired disc mass for each simulation.
We then tilt the planets and the two portions of the disc all by the same angle so that they are mutually coplanar but misaligned to the binary orbital plane. The initial surface density distribution for the simulations is the equilibrium for a coplanar disc.   The upper panels in Fig.~\ref{fig:splash} show the initial  set up for a simulation that is tilted by $40^\circ$ to the binary orbital plane. The upper middle panel shows that the two planets and the disc are all initially aligned with each other, but tilted with respect to the binary orbital plane.

\begin{figure*}
    \includegraphics[width=\columnwidth]{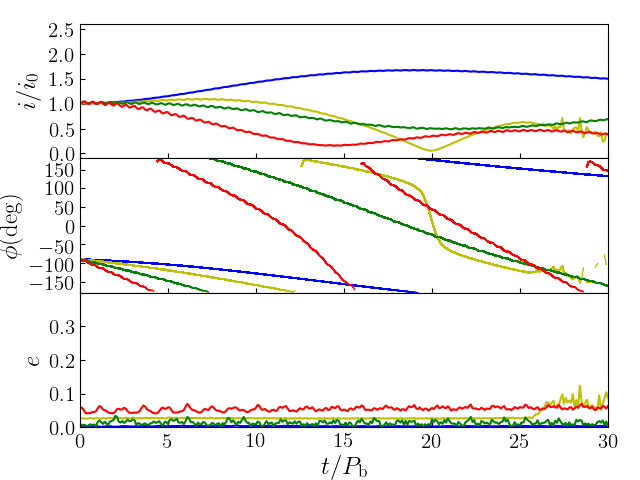}
    \includegraphics[width=\columnwidth]{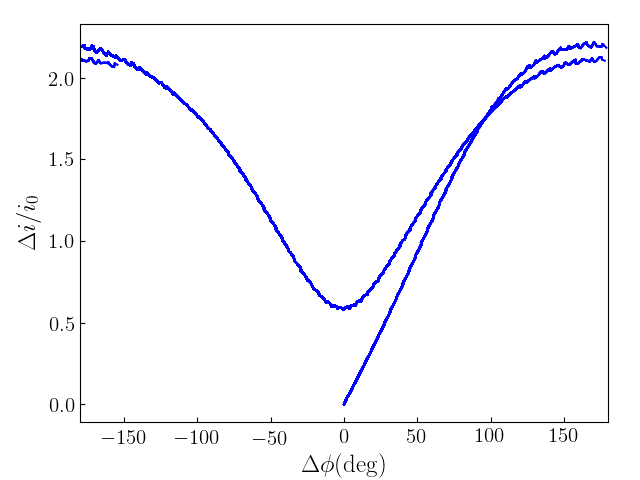}
    \caption{Left panel: Evolution of the inclination (upper panel), phase angle (middle panel) and eccentricity (bottom panel) of the components that orbit the primary star: inner planet (blue line), inner disc (yellow), outer planet (green) and outer disc (red) as a function of time in units of the binary orbital period $P_{\rm b}$ for a hydrodynamic simulation.
    Right panel: Phase portrait of the relative planet-planet tilt $\Delta i/i_0$ vs. nodal phase difference $\Delta \phi$.
    The system is initially coplanar, inclined by $i_0=10^{\circ}$ with respect to the binary plane and with total disc mass $M_{\rm d}=0.001\,M$.}
    \label{fig:sph_m001_i10}
\end{figure*}

\subsection{Low inclination system}

We first investigate the evolution of a system with initial total disc mass $M_{\rm d}=0.001\,M$ and  initial inclination $i=10^{\circ}$. The results are shown in Figure \ref{fig:sph_m001_i10}. We evaluate the inclination $i$ and phase angle $\phi$ for the inner and outer disc using a weighted density average from $R_{\rm in,1}=0.06\,a$ to $R_{\rm out,1}=0.125\,a$ and from $R_{\rm in,2}=0.175\,a$ to $R_{\rm out,2}=0.25\,a$ respectively. 

The two planets that orbit the primary are circulating with respect to one another, in agreement with the secular model with the same parameters. The inner disc (yellow) looks like it might be librating with the outer planet (green). However the inner disc is accreted after $30\,P_{\rm b}$, therefore it is hard to make predictions on their mutual evolution.  While in the analytical calculations we find the inner planet to be librating with the portion of the disc between the two planets, in our simulations this seems not to be the case and the two components are circulating.  The simulations suggest that the analytic model may have underestimated the gap size around the planets. However, the behaviour of the planets with respect to each other (circulating) is the same in both the analytic model and the simulations.  In the SPH simulations the gap size evolves depending on the relative inclination of the planet and the disc. This effect is not taken into account in the analytic model.
We show the results up to $30\,P_{\rm b}$ because at that time the disc between the planets is essentially accreted. We discuss the rapid accretion and how it affects our simulations further in Section~\ref{sec:concl}.


The right panel of Figure \ref{fig:sph_m001_i10} shows the phase portrait of the relative planet-planet tilt. We can see that this is consistent with the right panel of Figure \ref{fig:discs_m0.001} that shows that the two planets are circulating with each other. 
The increase in relative tilt between the two planets is in good agreement with the secular theory.

The critical disc mass required for the planets to librate is higher in the SPH simulation compared to the linear theory.
The reason is that in the simulations the disc is partially accreted onto the planets as the system evolves while the analytical calculations do not include accretion of mass.
Also, the surface density profile is assumed to be a power-law for both discs in the calculations in Section \ref{sec:script} while in the numerical simulations the profile is determined by the planets carving gaps in the disc. Furthermore the disc outer radius might be underestimated in the secular model because it does not take into account the viscous spreading of the disc.

\begin{figure}
    \centering
    \includegraphics[width=\columnwidth]{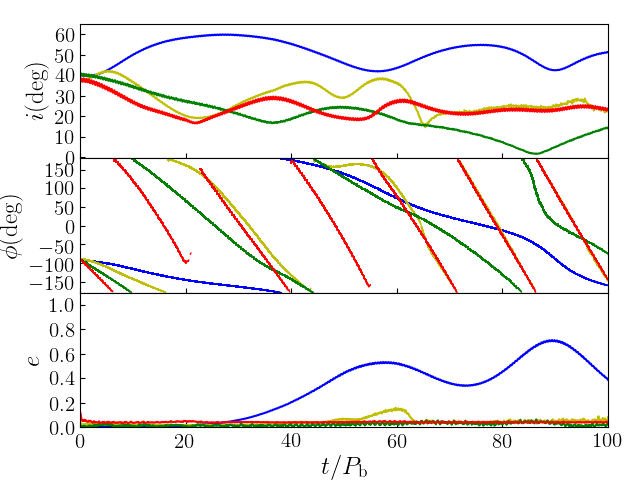}
    \caption{Same as Figure \ref{fig:sph_m001_i10} but with initial inclination of $i_0=40^{\circ}$ and with total disc mass $M_{\rm d}=0.001\,M$. Note that the yellow line is not distinguishable in the phase plot after roughly 50 orbits because the gap is filled and so the two parts of the disc behave in the same way.
    }
    \label{fig:sph_m001_i40}
\end{figure}

\begin{figure}
    \centering
    \includegraphics[width=\columnwidth]{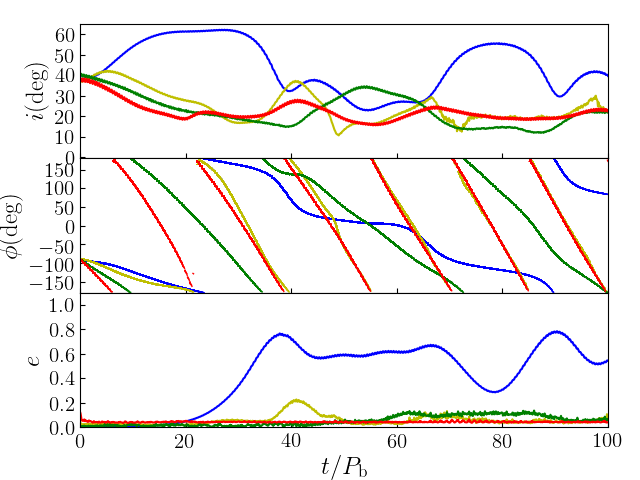}
    \caption{Same as Figure \ref{fig:sph_m001_i10} but with initial inclination of $i_0=40^{\circ}$ and with total disc mass $M_{\rm d}=0.002\,M$. Note that the yellow line is not distinguishable in the phase plot after roughly 50 orbits because the gap is filled and so the two parts of the disc behave in the same way.
    }
    \label{fig:sph_m002_i40}
\end{figure}

\begin{figure*}
    \includegraphics[width=\columnwidth]{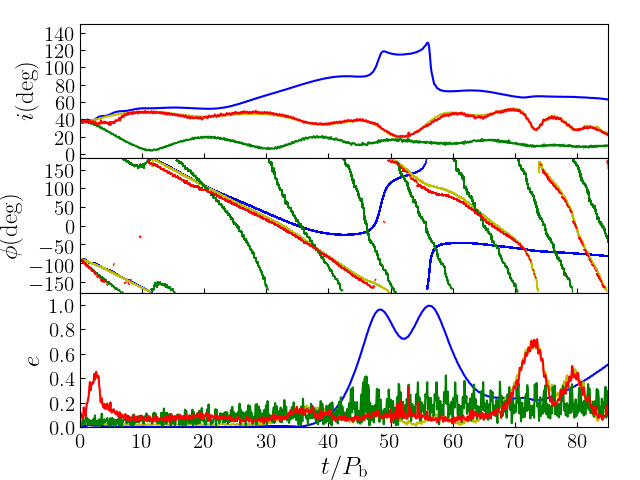}
    \includegraphics[width=\columnwidth]{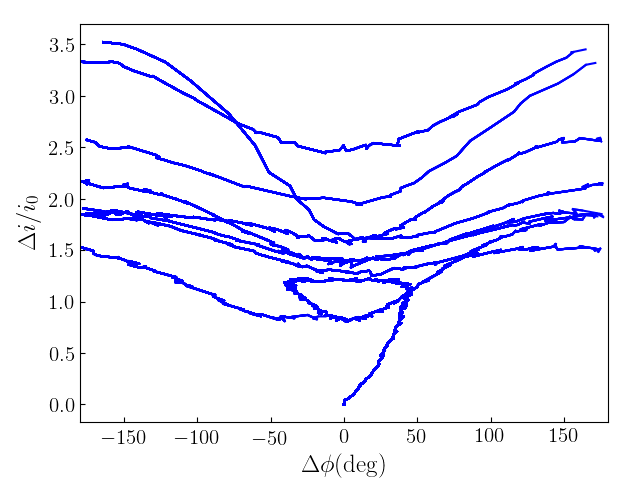}
    \caption{Left panel: same as Figure \ref{fig:sph_m002_i40} but with initial inclination of $i_0=40^{\circ}$ and with total disc mass $M_{\rm d}=0.02\,M$. Right panel: Phase portrait of the relative planet-planet tilt $\Delta i/i_0$ vs. nodal phase difference }
    \label{fig:sph_m02_i40}
\end{figure*}

\subsection{KL oscillations of the inner planet}

In order to explore the possibility of the inner planet undergoing eccentricity growth through KL oscillations  we increase the initial misalignment of the system with respect to the plane of the binary to $i=40^{\circ}$. If the initial inclination is higher, the portion of the disc between the two planets is likely to survive for longer since the misalignment between the planets and the disc becomes large enough to slow accretion of the disc material onto the planets. 

The evolution of the inclination, phase angle and eccentricity of the four components orbiting the primary star as a function of time are shown in Figure \ref{fig:sph_m001_i40} for initial total disc mass $M_{\rm d}=0.001\,M$ and in Figure \ref{fig:sph_m002_i40} for disc mass $M_{\rm d}=0.002\,M$.  For both simulations, the four components around the primary all undergo precession on different timescales initially. This can also been seen in the lower panels of Fig.~\ref{fig:splash} that show the higher mass disc at a time of $t=5\,P_{\rm orb}$. The inner and outer discs are clearly misaligned to each other and to the binary orbit. 
Later on in the simulation, the inclination of the outer planet to the disc becomes sufficiently  high that the torque from the outer planet is no longer strong enough to keep the gap open. Material flows through the gap and there is just one disc rather than two distinct components.

We can clearly see in Figs~\ref{fig:sph_m001_i40} and~\ref{fig:sph_m002_i40} that the inner planet, represented by the blue line, undergoes KL oscillations of eccentricity and inclination. For a lower mass disc, the inner planet reaches an inclination above the critical value on a slightly longer timescale and this results in a delay in the first KL oscillation.
Furthermore we can see that a higher disc mass results in larger amplitude KL oscillations. 

The KL oscillation period changes as the system evolves and we can see from Figure \ref{fig:sph_m002_i40} that the inner planet is starting a second KL cycle with a different periodicity at roughly $80\,P_{\rm b}$. The comparison between the results obtained using two different disc masses shows that the period of the first KL oscillation is shorter if the mass of the disc is larger, in agreement with the analytical  calculations that show shorter tilt oscillations periods for larger disc masses (see also right panel of Figure 2 in \cite{Lubow2016}).

Starting the multi--planet--disc system at a higher inclination results in libration between the two discs after roughly $60\,P_{\rm b}$ for the lower mass disc (Fig. \ref{fig:sph_m001_i40}) and about $20\,P_{\rm b}$ for the higher mass disc (Fig. \ref{fig:sph_m002_i40}). The misalignment between the outer planet and the two discs becomes large enough so that the planet torque is weaker and the disc material can replenish the gap carved by the outer planet.
In the  bottom panel in Fig. \ref{fig:sph_m001_i40}), the disc between the planets starts to librate with the inner planet and becomes slightly eccentric at roughly $60\,P_{\rm b}$. Then the disc material replenishes the gap and the inner disc starts to librate, as it circularizes, with the outer disc. 
The same mechanism occurs on a shorter timescale for the more massive case (Fig. \ref{fig:sph_m002_i40}). 

We can see from Figures \ref{fig:sph_m001_i40} and \ref{fig:sph_m002_i40} that if the two planets are circulating the inner planet undergoes KL oscillations while the outer planet remains on a circular orbit and does not affect the evolution of the inner one.

We further increase the mass of the disc to $M_{\rm d}=0.02\,M$ in order to understand the role of the second planet on the KL oscillations of the inner planet. 
The evolution might be significantly different if the disc is massive enough for the two planets to start librating together. In this scenario the outer planet could cause damping of the KL oscillations of the inner planet. 

Figure \ref{fig:sph_m02_i40} shows the evolution of the inclination, phase angle and eccentricity (left panel) of each component that orbits the primary star and the phase portrait (right panel) of the two planets. 
We can see from the right panel that the two planets start to librate at the very beginning of the simulation when the mass of the disc is high enough. 
Furthermore since the mutual inclination between the outer planet and the two discs increases very rapidly with time, the gap is replenished and therefore the disc starts to behave as a whole and undergoes two KL oscillations after roughly $70\,P_{\rm b}$.

The higher mass disc does not seem to affect the possibility of having KL eccentricity growth of the inner planet, infact, it strengthens its effect. The inner planet reaches a much higher eccentricity and inclination value compared to the lower disc mass cases.
We can see from the left panel of Figure \ref{fig:sph_m02_i40} that the inner planet inclination increases beyond $90^{\circ}$, meaning that the planet is orbiting in a retrograde fashion before coming back to the prograde orbit.

\begin{figure}
    \centering
    \includegraphics[width=\columnwidth]{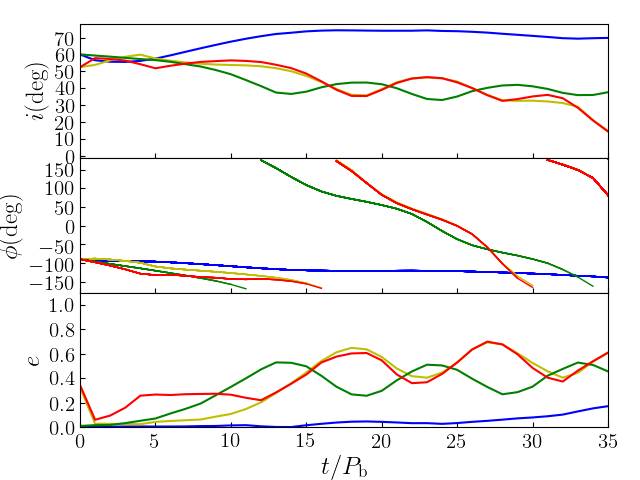}
    \caption{Same as Figure \ref{fig:sph_m001_i40} but starting at $i_0=60^{\circ}$.}
    \label{fig:sph_m001_i60}
\end{figure}

\subsection{KL oscillations of the outer planet}

We have also investigated the possibility for the outer planet to undergo KL oscillations.
From Figs. \ref{fig:sph_m001_i40}-\ref{fig:sph_m02_i40} we can see that the outer planet inclination decreases dramatically during the first tens of binary orbits. The higher the disc mass, the faster this decrease towards $i=0^{\circ}$ occurs. 
Therefore the system would have to start at higher inclinations than those explored so far in order for the outer planet inclination to exceed the critical angle for the KL mechanism to operate.

We ran a simulation with a disc mass of $M_{\rm d}=0.001\,M$ starting at a $60^{\circ}$ inclination. The results are shown in Figure \ref{fig:sph_m001_i60}.
The outer planet is indeed able to undergo eccentricity and inclination oscillations and to maintain a moderate eccentricity ($e\approx0.5$) for the whole duration of the simulation.
At the same time the two parts of the disc undergo KL oscillations as well since the initial inclination is well above the critical angle \citep{Martin2014,Fu2015,Lubow2017,Franchini2019}.
Since the discs become very eccentric they accrete more efficiently onto the central star and are completely drained after $40\,P_{\rm b}$.

\begin{figure*}
    \centering
    \includegraphics[width=2\columnwidth]{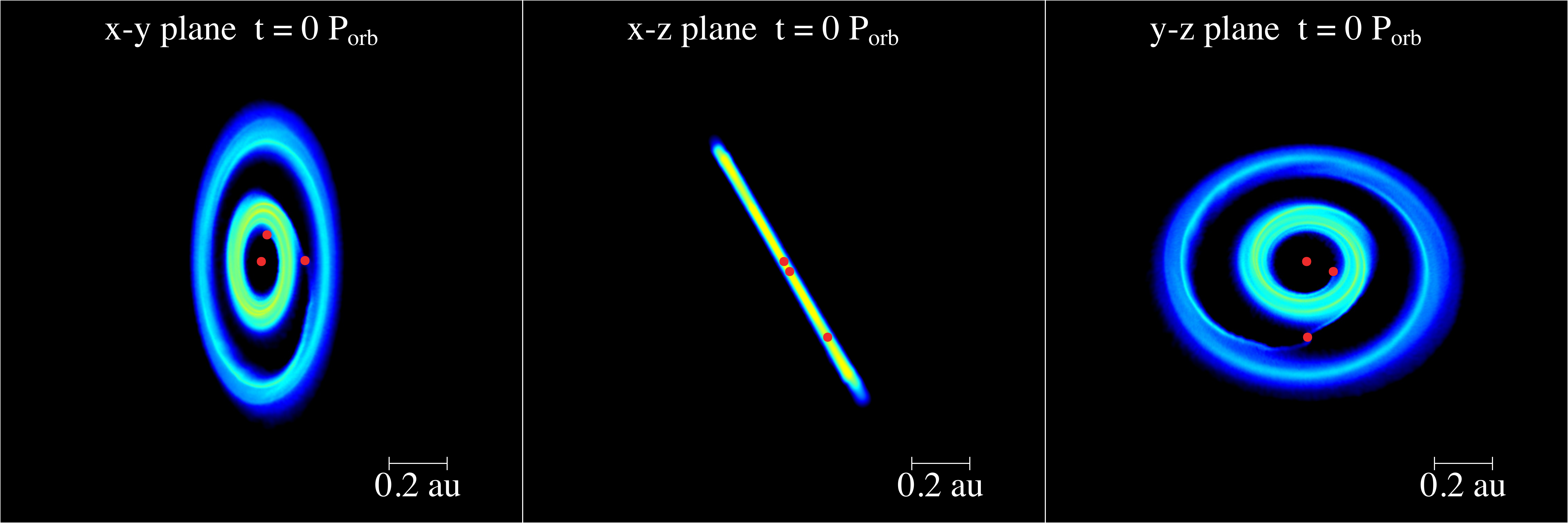}
     \includegraphics[width=2\columnwidth]{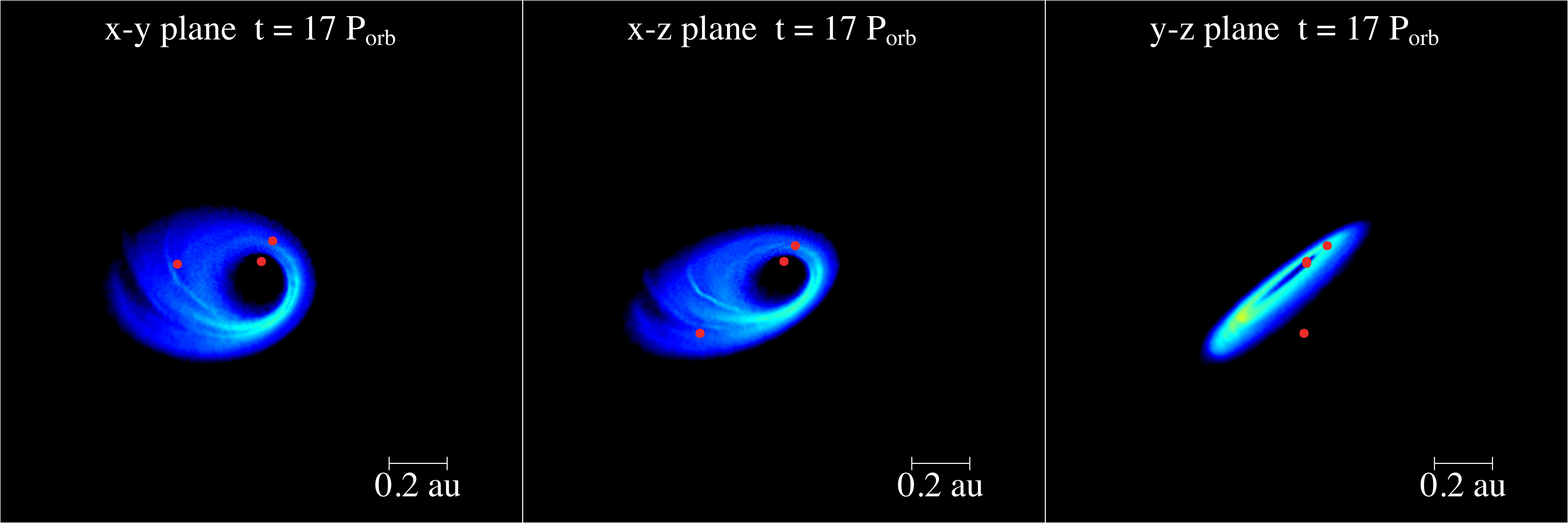}
    \caption{ Same as Fig.~\ref{fig:splash} except for a simulation with initial disc mass $M_{\rm d}=0.001\,M$ and tilt $60^\circ$. The upper panels show time $t=0$ while the lower panels show time $t=17\,P_{\rm orb}$.  }
    \label{fig:splash2}
\end{figure*}

 Fig.~\ref{fig:splash2} shows the  disc-binary-planet  at the start of the simulation and at a time $t=17\,P_{\rm orb}$.  The outer planet is sufficiently misaligned to the disc that material flows through the planet gap. The disc itself is seen to be highly eccentric as a result of its KL oscillations.

\section{Conclusions}
\label{sec:concl}

In this paper we investigated the interaction between  two  giant planets embedded in a protoplanetary disc that orbits one component of a binary system.
The planets and the disc are initially coplanar but misaligned with respect to the binary orbital plane.
We first investigated the effect of the disc in determining whether the planets librate together or precess independently (i.e. circulate) using the secular model \citep{Lubow2016}. The results show that the presence of the disc leads to larger mutual inclinations between the two planets. Furthermore, for less massive discs ($M_{\rm d}\leq0.001\,M$) the planets circulate while for higher disc masses the planets precess together around the primary star.

Since the secular model assumes small tilts and its results are therefore accurate up to about $20^{\circ}$, we then used SPH simulations to investigate the possibility of KL oscillations of the planets starting with the planets and the disc misaligned by $i=40^{\circ}$.
In particular we find that the inner and outer planets, as well as the disc, can undergo eccentricity growth depending on the initial inclination of the planets-disc system and the disc mass.
Furthermore, more massive discs lead to larger oscillation amplitudes of the inner planet eccentricity and inclination.

We confirmed the prediction of the secular model with our numerical simulations in terms of how the disc influences the planets behaviour.
However the critical disc mass for the planets to librate together is very different in the two approaches because of the presence of viscosity and therefore accretion of mass  onto the planets in the SPH simulations.
If the disc is as massive as $M_{\rm d}=0.02\,M$, we find that the two planets do start to librate at the very beginning of the simulations but they switch to circulation when the disc mass decreases from the initial value.
The inner planet reaches an eccentricity close to unity and transitions to a retrograde orbit before its oscillations are damped and it returns to orbit the primary star prograde.
Therefore the interaction between two planets and the massive disc where they are embedded might explain the formation of  observed retrograde planets such as WASP-17b \citep{Anderson2010}.

 One limitation of our simulations is that the disc disperses on a short timescale.  This is a consequence of several factors. First, the accretion radii of the planets and the central star cause material to accrete faster than it otherwise should. Secondly, there may be some overlap in the planet gaps that would cause the disc between the two planets to be depleted on a short timescale \citep[e.g.,][]{Morbidelli2007}. This may not be an important effect in the case of the highly misaligned systems that we have considered here. A more massive disc would result in  larger tilt oscillations and a wider range of conditions for mutual libration between the planets than we found here. For the types of systems we considered, a higher disc mass would also mean that the inner planet would be more likely to undergo KL oscillations while the outer planet would be less likely.   

Note that if we want to further increase the mass of the accretion disc we would need to consider the effect of the disc self-gravity.  \cite{Fu2015b} showed for an equal mass binary that the disc mass required to quench KL disc oscillations is a few percent of the mass of the central star. Therefore we expect that if we increase the mass of the disc up to, say, $M_{\rm d}=0.05\,M$ the KL oscillations of the disc are likely to be suppressed. 

\section*{Acknowledgements}

We thank Giovanni Dipierro for useful comments and discussion.
We thank Daniel Price for providing the {\sc phantom} code for SPH simulations and acknowledge the use of {\sc splash} \citep{Price2007} for the rendering of the figures. We acknowledge support from NASA through grant NNX17AB96G.  Computer support was provided by UNLV's National Supercomputing Center.










\bsp	
\label{lastpage}
\end{document}